\documentclass{emulateapj}
\def\geqsim{\lower.73ex\hbox{$\sim$}\llap{\raise.4ex\hbox{$>$}}$\,$}
\def\leqsim{\lower.73ex\hbox{$\sim$}\llap{\raise.4ex\hbox{$<$}}$\,$}

\def\Mpch {~h^{-1}~{\rm Mpc}}
\def\kpch {~h^{-1}~{\rm kpc}}
\def\kms{~{\rm km~s^{-1}}}
\newcommand{\Hii}{\ion{H}{2}}

\newcommand{\myemail}{admyers@astro.uiuc.edu}
\newcommand{\beq}{\begin{equation}}
\newcommand{\eeq}{\end{equation}}
\newcommand{\MgII}{Mg\,{\sc II}}
\newcommand{\CaII}{Ca\,{\sc II}}
\newcommand{\CIV}{C\,{\sc IV}}
\newcommand{\CIII}{C\,{\sc III}]}

\slugcomment{ApJ, in prep \today}
\shorttitle{}
\shortauthors{Myers et al.}

\begin{document}

\title{Quasar Clustering at $25\kpch$ from a Complete Sample of Binaries\altaffilmark{1}}

\author{Adam D. Myers\altaffilmark{2}, Gordon T. Richards\altaffilmark{3}, Robert J. Brunner\altaffilmark{2,4}, Donald P. Schneider\altaffilmark{5}, Natalie E. Strand\altaffilmark{6}, Patrick B. Hall\altaffilmark{7}, Jeffrey A. Blomquist\altaffilmark{3}, and Donald G. York\altaffilmark{8}}

\email{\myemail}

\altaffiltext{1}{Some data presented here were obtained at Kitt Peak National Observatory, a division of the National Optical Astronomy Observatories, which is operated by the Association of Universities for Research in Astronomy, Inc. under cooperative agreement with the National Science Foundation.}
\altaffiltext{2}{Department of Astronomy, University of Illinois at Urbana-Champaign,ÊUrbana, IL 61801}
\altaffiltext{3}{Department of Physics, Drexel University, 3141 Chestnut Street, Philadelphia, PA 19104}
\altaffiltext{4}{National Center for Supercomputing Applications, Champaign, IL 61820}
\altaffiltext{5}{Department of Astronomy and Astrophysics, 525 Davey Laboratory, Pennsylvania State University, University Park, PA 16802}
\altaffiltext{6}{Department of Physics, University of Illinois at Urbana-Champaign, Urbana, IL 61801}
\altaffiltext{7}{Department of Physics and Astronomy, York University, Toronto, ON M3J 1P3, Canada}
\altaffiltext{8}{Department of Astronomy and Astrophysics, University of Chicago, Chicago, IL 60637}

\begin{abstract}

We present spectroscopy of binary quasar candidates selected from Data Release 4 of the Sloan Digital Sky Survey (SDSS DR4) using Kernel Density Estimation (KDE). We present 27 new sets of observations, 10 of which are binary quasars, roughly doubling the number of known $g < 21$ binaries with component separations of $3\arcsec\leq\Delta\theta<6\arcsec$. Only 3 of 49 spectroscopically identified objects are non-quasars, confirming that the quasar selection efficiency of the KDE technique is $\sim95$\%. Several of our observed binaries are wide-separation lens candidates that merit additional higher-resolution observations. One interesting pair may be an M star binary, or an M star-binary quasar superposition. Our candidates are initially selected by UV-excess ($u-g < 1$), but are otherwise selected irrespective of the relative colors of the quasar pair, and we thus use them to suggest optimal color similarity and photometric redshift approaches for targeting binary quasars, or projected quasar pairs. From a sample that is complete on proper scales of $23.7 < R_{prop} < 29.7\kpch$, we determine the projected quasar correlation function to be $\overline{W_p}=24.0 \pm^{16.9}_{10.8}$, which is $2\sigma$ lower than recent estimates. We argue that our low $\overline{W_p}$ estimates may indicate redshift evolution in the quasar correlation function from $z\sim1.9$ to $z\sim1.4$ on scales of $R_{prop} \sim25\kpch$. The size of this evolution broadly tracks quasar clustering on larger scales, consistent with merger-driven models of quasar origin. Although our sample alone is insufficient to detect evolution in quasar clustering on small scales, an $i$-selected DR6 KDE quasar catalog, which will contain several hundred $z~\leqsim~5$ binary quasars, could easily constrain any clustering evolution at $R_{prop} \sim25\kpch$.
\end{abstract}


\section{Introduction}

Cosmologically, quasars can now be explained as one spectacular stage of an evolutionary process, initiated by gas-rich galaxy mergers, that ultimately helps redden elliptical galaxies (see, e.g., \citealt{Hop06,Hop07}). That quasar activity might trace the early stages of merger-driven galaxy evolution makes quasar observations an essential ingredient in constraining galaxy formation scenarios. On the other hand, less luminous Active Galactic Nuclei (AGN), particularly at low redshift ($z~\leqsim~1$), may be better explained by less violent fueling mechanisms (e.g., \citealt{HandH06}) than the mergers that drive quasars' optical intensity. Given that orientation can complementarily explain many differences between the AGN zoo (e.g., \citealt{Ant93,Elv00}), new AGN constraints over a broad range of luminosity, particularly across $z\sim1$, may be key to determining which elements of quasar behavior are mainly structural, and which are mainly evolutionary.

If quasars are associated with galaxy mergers, observations of binary quasars with proper separations comparable to the scale of small galaxy groups should offer interesting constraints. It has generally become accepted that most quasar pairs at similar redshifts that have image separations $\geqsim~3\arcsec$, are binary quasars rather than lenses \citep{Phi86,Bah86,Koc99,Rus02,Ogu06}. AGN activity can be exacerbated by tidal forces in galaxy mergers \citep{Bar96,Bah97} and it has long been argued that this might explain an excess of binary quasars \citep{Djo91,Koc99,Mor99} and physical triples \citep{Djo07}. \citet{Hop07pre} suggest instead that a binary quasar excess simply reflects the increased probability for mergers to occur in regions that are overdense on small scales. If quasars form in mergers they will thus be naturally more biased at small scales. \citet{Hop07pre} further argue that orbits for which quasar activity might be exacerbated in both of two merging galaxies are prohibitively rare, even if few such events are needed to explain a binary quasar excess (\citealt{Mye07b}; henceforth M07b).

The Sloan Digital Sky Survey (henceforth SDSS; \citealt{Yor00}) has renewed interest in binary quasars, and \citeauthor{Hen06} (\citeyear{Hen06}; henceforth H06) have used SDSS data to categorically confirm earlier evidence (e.g., \citealt{Djo91,Hew98}) that quasar clustering is enhanced on comoving scales $\leqsim~100\kpch$ (proper scales of $\leqsim~40\kpch$ at $z\sim1.5$). Binary quasars are scarce, with perhaps fewer than five hundred in the entire sky to $g~\leqsim~21$, redshifts of $z~\leqsim~2.5$, and comoving separations $\leqsim~100\kpch$ (given the sample size in M07b). Deep, wide imaging, such as from the SDSS, is thus key to testing predictions of the nature of binary quasars; for instance, by studying quasar clustering as a function of redshift or luminosity. If enhanced small-scale quasar clustering is due to the enhanced bias of major galaxy mergers, rather than tidal forces exacerbating quasar activity, then there should be no redshift evolution in the relative bias of quasar clustering on large and small scales. Further, some merger-driven models predict stronger small-scale clustering for quasars than for lower-luminosity AGNs, due to different fueling mechanisms (e.g., \citealt{Hop07pre}).

Binary quasar samples are typically assembled from pairs that are initially targeted as possible gravitational lenses (e.g., \citealt{Mor99}; H06). Because of this, the two members of most known binary quasars have similar colors. Although color similarity may be optimal in detecting lenses, scatter in the quasar color-redshift relation (e.g., \citealt{Ric01}) dictates that strict color similarity cannot select {\em all} binary quasars. As binary quasars are useful in testing merger-driven models of quasar activity, it is disconcerting that color similarity cuts might discard particularly informative binaries, such as any that are being exacerbated by tidal forces in merging galaxies. Further, as binary quasars are scarce, relaxing color criterion might help provide enough binaries to study their redshift evolution.

We have now performed spectroscopy of a sample drawn from a complete, UVX ($u-g < 1$) set of SDSS Data Release 4 (henceforth DR4; \citealt{Ade06}) binary quasar candidates (see Table~1 of M07b). The sample, $\sim45\%$ of which has now been identified, is photometrically selected using the Kernel Density Estimation (KDE) technique of \citet{Ric04}. Our aim is to compile an extensive, homogeneous set of binary quasars, with proper separations $\leqsim~40\kpch$, mainly to study quasar clustering on small scales, ultimately as a function of redshift. Most of our binary candidates are unlikely to be observed in the main SDSS quasar survey (e.g., \citealt{Sch07}), because our observations probe fainter, and because SDSS fibers (in single tiles) cannot be placed closer than 55$\arcsec$. Our approach differs from previous studies of binary quasars. In particular, our main goal is to study binary quasars, not gravitational lenses (unlike the samples compiled in, e.g., \citealt{Koc99,Mor99}; H06). Thus, excepting our initial UVX cut, our binaries are the first sample selected regardless of the relative colors of the component quasars. This allows an investigation of whether color similarity can be used to optimally target binary quasars (\S\ref{sec:colprop}). 

In \S\ref{sec:data}, we discuss our initial results in compiling a homogeneous, spectroscopic binary quasar sample from the SDSS, and report 27 sets of new observations of binary quasar candidates. In \S\ref{sec:clus} we use a subset of these observations to present the first analysis of quasar clustering on small scales using a {\it complete} spectroscopic sample of binary quasars. All our binaries are selected from the DR4 KDE candidates of M07b, and are therefore a straightforward subset of SDSS DR4, making our selection function very simple. We correct all magnitudes for Galactic extinction using the maps of \citet{Sch98}, unless otherwise noted. We adopt $\Omega_m=0.27$, $\Omega_\Lambda=0.73$, consistent with WMAP3 \citep{Spe07}, and $h\equiv H_0/100{\rm km~s^{-1}~Mpc^{-1}} = 0.7$. We denote transverse proper (comoving) scales as $R_{prop}$ ($R$) and radial proper (comoving) scales as $s_{prop}$ ($s$).

\section{Data}
\label{sec:data}

\subsection{Observations}

\subsubsection{Candidate Selection}

Our candidate binary quasars are photometric objects in SDSS DR4 that are classified as quasars by the KDE technique \citep{Ric04}, have $u-g < 1$, $g < 21$, and are within 3\arcsec~to 6\arcsec~of another such object. The SDSS $ugriz$ filters are described in \citet{Fuk96}. As in M07b, we inspect our candidates and discard any that are clearly not quasars (close KDE candidates are occasionally misclassified \Hii{} regions in low-redshift galaxies). We restrict our spectroscopy, and analysis, to quasar pairs with angular separations of $\Delta\theta > 3\arcsec$, so that components of a pair appear clearly separated in SDSS imaging (e.g., H06). Although scales that are comparable to small galaxy groups or smaller are most useful in testing merger models of quasar activity, our upper limit of $\Delta\theta < 6\arcsec$ is somewhat arbitrary, providing a reasonable number of candidates for a small spectroscopic study. Table~1 of M07b lists our 98 DR4 candidate binaries (and a further 13 candidates with $\Delta\theta < 3\arcsec$). At $g < 21$ the KDE technique, coupled with a UVX ($u-g < 1$) cut, selects quasars with 95\% efficiency, and is over $95\%$ complete for redshifts of $0.2~\leqsim~z~\leqsim~2.4$ (e.g., \citealt{Ric04,Mye06,Mye07a}). Our survey goal is an efficiently classified, statistical, sample of binary quasars, which are UVX but are otherwise selected irrespective of the relative colors of the components of the binary. At $z~\leqsim~2.5$, our selection should thus only bias our sample against those binaries where one or both component quasars is reddened beyond $u-g=1$.

\subsubsection{Spectroscopy}

Spectroscopy of our DR4 KDE binary quasar candidates was obtained with the R-C Spectrograph on the Mayall 4-m, over 5 nights (UT 2007 February 22--26) at Kitt Peak National Observatory. We used a $1.5\arcsec$ by $98\arcsec$ long-slit set at the position angle of the candidate binary, allowing both components to be simultaneously observed. The KPC-10A grating and WG360 blocking filter yielded a resolution of $\sim5$\AA~and a wavelength coverage of $\sim3800$--7500\AA. The most useful observations were obtained on February 25 and 26, due to cloud and wind on other nights. Over February 25 and 26, the seeing was $\leqsim~1.5$\arcsec, allowing even our closest binary candidates ($\sim$3\arcsec) to be spatially separated. The survey goals were a positive identification and redshift for each DR4 KDE binary candidate. This typically required a 15~minute exposure when the {\it faintest} member of the candidate binary was at $g\sim19.5$ and three 20~minute exposures (or longer) for $g\sim21$, although the $\geqsim~70$\% illuminated Moon on February 25 and 26 typically prevented our  $g\sim21$ candidates from being spectroscopically identified.

Spectra were reduced at the telescope, using IRAF\footnote{Distributed by the National Optical Astronomy Observatory, which is operated by the Association of Universities for Research in Astronomy, Inc., under cooperative agreement with the National Science Foundation.}. Exposures ceased once a binary candidate could be identified as; (1) containing one star or galaxy; or (2) containing two quasars with established redshifts. We estimate redshifts using the rest frame emission wavelengths listed in Table~2 of \citet{Van01}. Based on deriving redshifts, where possible, from several different lines in each quasar's spectrum, our typical error is $\Delta z\sim0.0031$. A similar approach suggests a mean velocity precision of $\sim370\kms$, with little redshift dependence. We note that this is likely an overestimate, as velocity differences between quasars are usually found to be more precise when cross-correlating the full spectra, rather than measuring line shifts (e.g., \citealt{Ton79,Djo84}). 

Tables~\ref{table:singles}--\ref{table:ambig} detail our new observations. Table~\ref{table:singles} lists objects for which we obtained an identification for only one member of the candidate binary. Table~\ref{table:bin} lists confirmed binary quasars.  Following H06, we classify quasar pairs with a line-of-sight velocity difference of $|\Delta v_{\|}|<2000\kms$ ($s_{prop} < 9.5\Mpch$ at $\bar{z}=1.4$) as a binary. Table~\ref{table:nonbin} lists DR4 KDE binary candidates that are actually ``projected'' quasar pairs (with components that lie at different redshifts), or pairs in which one object is not a quasar. When both redshifts listed in Table~\ref{table:nonbin} have some uncertainty, a binary quasar interpretation can still be ruled out, based on strong lines observed in both quasars but at discrepant wavelengths. Pairs that are harder to definitively identify are listed in Table~\ref{table:ambig} (see \S\ref{sec:IntObj}). 

In all, we spectroscopically identified 49 objects, mainly in areas with Galactic absorption $A_g~\leqsim~0.17$. Of these identified objects, 44 are both members of 22 candidate quasar pairs, and 5 are objects from pairs for which we identified only one member. Of the 49 identified objects 46 are quasars, confirming the KDE technique is $\sim95\%$ efficient for $A_g~\leqsim~0.21$ \citep{Mye07a}. Of the 22 candidate quasar pairs for which we identified both components, 3 are quasar-non-quasar pairs, 9 are projected quasar pairs (i.e. at disjoint redshifts), and 10 are binary quasars. Several of the 10 binary quasars could, in fact, be previously unrecorded lenses; see \S\ref{sec:Lens}). In Table~\ref{table:old} we list previously known quasar pairs that also meet the criteria to be included in our DR4 KDE binary candidate sample. Previously known candidates include 12 non-binaries, 9 binaries, and 2 lenses. Approximately half of the 98 ($3\arcsec\leq\Delta\theta<6\arcsec$) DR4 KDE candidates have now been spectroscopically identified (see Table~\ref{table:bkdown}). With the caveat that bright objects may have been observed first, $\sim42\%$ of the candidate pairs are binary quasars, and only $\sim16\%$ of the pairs contain a non-quasar.

\subsection{Interesting Spectroscopic Pairs} 

\subsubsection{Potential Lenses}

\label{sec:Lens}

\begin{figure}
\plotone{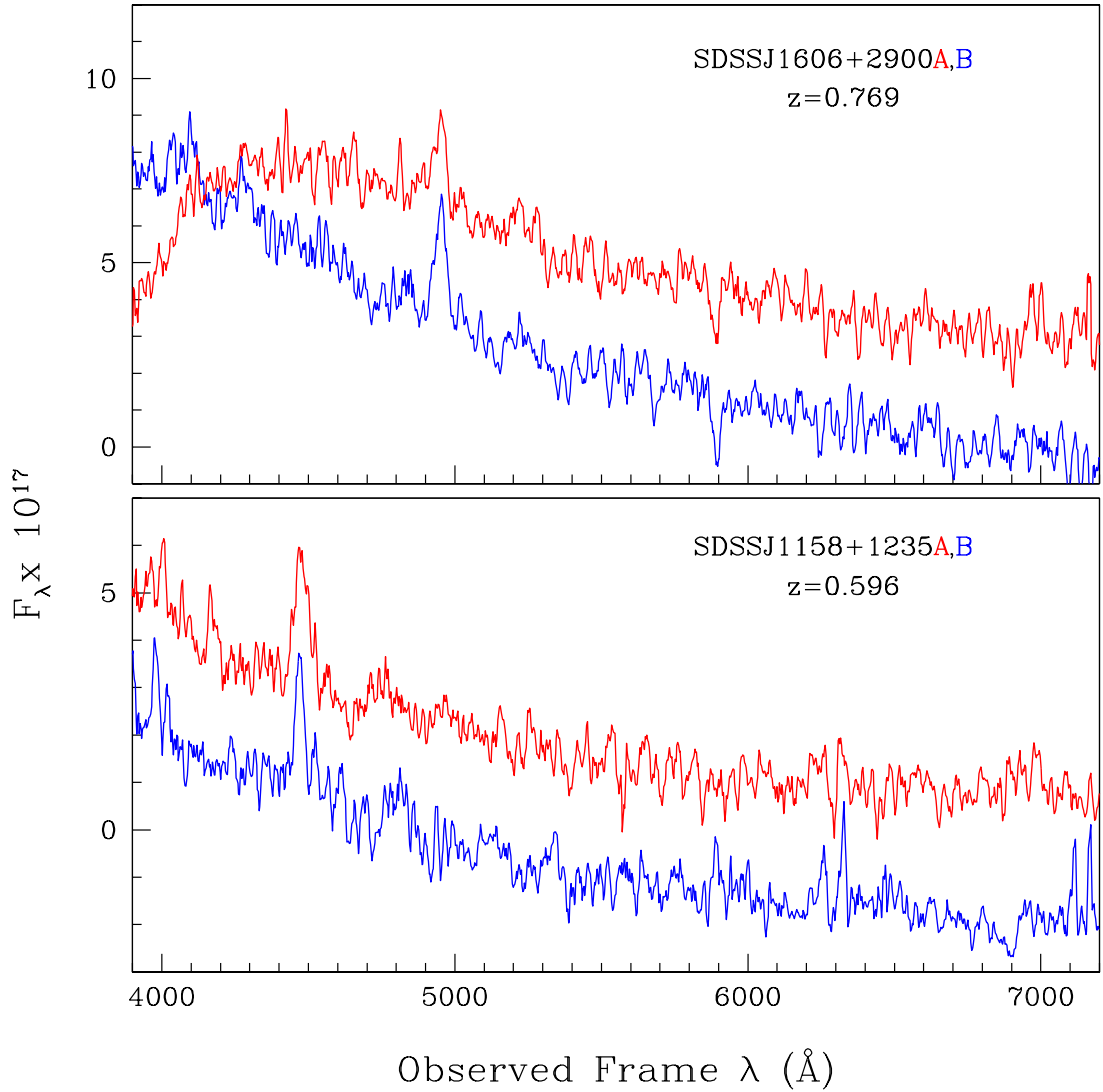}
\caption{Two of the five quasar pairs in our DR4 KDE sample that require higher resolution spectroscopy to determine whether they are binary quasars or a lensed quasar. These spectra were taken with the R-C Spectrograph on the Mayall 4-m at KPNO at a resolution $\sim5$\AA, and have been smoothed with a 5~pixel boxcar. SDSSJ1158+1235A,B and SDSSJ1606+2900A,B are our two most likely lens candidates, based on their similar colors and component separations of $\Delta\theta < 4\arcsec$. In both panels, component A is the upper spectrum and component B has been offset, which can cause component B to have a flux density below zero. In each case, both components are, in reality, at nearly identical flux levels.
\label{fig:Lens}}
\end{figure}

Five pairs in Table~\ref{table:bin} have sufficiently similar spectra, at our $\sim5$\AA~resolution, that they might be a lensed quasar rather than a binary. As lenses with image separations in the range $3\arcsec\leq\Delta\theta<6\arcsec$ are rare (e.g., \citealt{Ina07}), particularly for $z~\leqsim~2$, we interpret these objects as binaries (see also \citealt{Koc99}), although they certainly merit higher-resolution spectroscopy. A lensing interpretation is especially unlikely for the three possible lenses used in our clustering analysis (\S\ref{sec:clus}), which have $\sim5\arcsec$ separations and, in two cases, dissimilar colors. In Figure~\ref{fig:Lens} we display the spectra of our most likely lens candidates. SDSSJ1158+1235A and B, in particular, have almost identical spectra at our resolution.

\subsubsection{Notes on Ambiguous Binaries}
\label{sec:IntObj}

Table~\ref{table:ambig}, lists four candidates that we could not definitively identify. We conclude that two of these objects are binaries, for the following reasons, quoting all wavelengths in the observed frame.

SDSSJ093424.32+421130.8 and SDSSJ093424.11+421135.0 consist of a quasar at $z=1.339$ and a featureless spectrum (after 4200s of exposure). SDSSJ093424.11+421135.0 is faint (observed $g=21.01$), probably a star, and has no obvious emission near 4460\AA~or 6550\AA, the principal emission lines used to identify SDSSJ093424.32+421130.8. We therefore conservatively conclude that this is not a binary quasar.

SDSSJ120727.09+140817.1 and SDSSJ120727.25+140820.3 both have ambiguous redshifts. The fainter object (observed $g=20.39$) has broad emission at 4320\AA,~and near 7840\AA~at the red edge of our coverage. Although the brighter object ($g=20.27$), has possible, low signal-to-noise ratio, emission near 4340\AA~we discount it based on the 4320\AA~emission in the fainter object being strong, and that the objects are, obviously, observed under similar conditions. We tentatively conclude that this pair is not a binary quasar.

\begin{figure}
\plotone{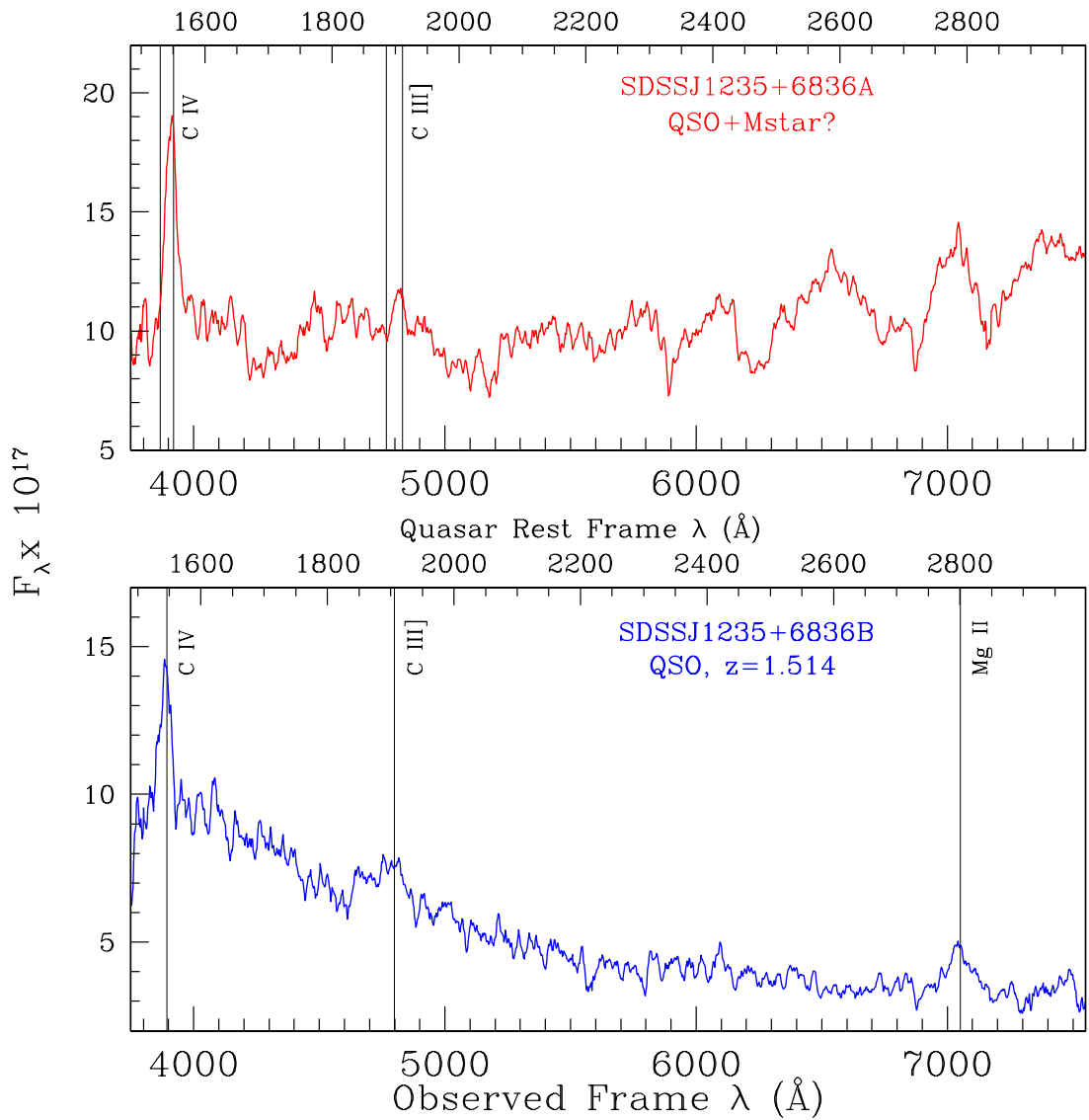}
\caption{Spectra of SDSSJ1235+6836A,B taken with the R-C Spectrograph on the Mayall 4-m at KPNO at a resolution $\sim5$\AA, and smoothed with a 5~pixel boxcar. In the lower panel, the vertical lines mark common quasar emission lines. For component A, in the upper panel, emission lines are marked at the systemic redshift of {\em component B} offset by $\pm2000\kms$, a typical window for a binary quasar (e.g., H06). The systemic redshift of component A in Table~\ref{table:ambig} is derived assuming the emission line near 3900\AA~is \CIV, and is close to the red side of these windows ($z=1.529$). At red wavelengths, component A resembles an M star.
\label{fig:ambig}}
\end{figure}

SDSSJ1235+6836A,B (see Figure~\ref{fig:ambig}) are an interesting pair with highly dissimilar colors. SDSSJ1235+6836B is apparently a quasar with significant broad emission near 3895\AA~and 7050\AA~and weaker emission near 4790\AA.  SDSSJ1235+6836B is most likely a quasar, with probable broad emission near 3910\AA~and possible emission near 4820\AA, lying behind a classic M star observed at the red end of the spectrum. As $\leqsim~3900$\AA, is near the blue edge of our coverage, an alternative possibility is that one or both objects are just M stars with very strong, blended \CaII~H and K emission. Assuming the proper motion of the M star(s) is moderate, this object may look more or less like an M star-quasar pair over time.

SDSSJ1507+2903A,B have strong emission near 5250\AA~and 5215\AA, respectively. Neither spectrum has additional features over 3800--7500\AA, and so we assume that the emission is \MgII, placing both quasars at $z\sim0.87$. The ambiguity for this pair is that their redshifts imply $|\Delta v_{\|}|=2100\kms$. As a shift of $\delta z < 0.0005$, far smaller than our typical precision, can bring these quasars within $|\Delta v_{\|}|<2000\kms$, we identify this pair as a binary.

\section{Color Selection of Binary Quasars}
\label{sec:colprop}

To study relative color selection of binary quasars, we use the $\chi^2$ color similarity statistic introduced by H06

\beq
\chi_{color}^2(A) = \sum_{\rm ugriz}\frac{(f^i_2- A f^i_1)^2}
    {[\sigma^i_2]^2 + A^2 [\sigma^i_1]^2} \label{eqn:chi}
\eeq

\noindent where the subscripts 1 and 2 represent the components of a pair. The superscript $i$ refers to flux ($f$) in the 5 SDSS bands ($ugriz$). For {\it asinh} magnitudes ($m$)

\begin{eqnarray}
  f^i &=& 2 F_0 b^i \sinh \left [ -m^i/P - \ln b^i \right] \nonumber \\ 
 \sigma_f^i &=& (\sigma_m^i/P)\sqrt{(2F_0b^i)^2+(f^i)^2}
  \label{eq:mag2flux}
\end{eqnarray}

\noindent where $P= 2.5/\ln10$ \citep{Pog56}, $F_0 = 3630.78$Jy, and 
$b^{[u,g,r,i,z]} = [1.4 , 0.9 , 1.2 , 1.8 , 7.4 ] \times 10^{-10}$ \citep{Lup99,Sto02}.  A quasar pair with more similar colors has a lower $\chi_{color}^2$. Iterative equations for calculating $A$ in Equation~\ref{eqn:chi} can be ill-conditioned for $\chi_{color}^2~\geqsim~30$, so, throughout this work, we numerically determine $A$ by bisection.

Binary quasars are often pairs rejected from gravitational lens searches, and as such, the components of known binaries typically have very similar colors, a long-known example being SDSSJ1637+2636A,B ($\chi_{color}^2=2.8$; see Table~\ref{table:old}; \citealt{Sra78,Djo84}). Schemes designed to optimize binary quasar searches by selecting pairs with similar colors, will, therefore, naturally reselect known binary quasars. Our DR4 KDE objects are simply all candidates with a high probability of being quasars and thus, after the initial homogeneous UVX cut, are selected irrespective of the {\it relative} colors of the components of the pair. The UVX cut itself, at $z~\leqsim~2.5$, should only bias our sample against those binaries with a component that is intrinsically dust-reddened beyond $u-g=1$. Our sample should thus be useful in determining color similarity cuts to optimize binary quasar selection. However, some of the quasar pairs in Table~\ref{table:old} were selected by H06 to have $\chi_{color}^2 < 20$; as we avoided reobserving these pairs, our data in Tables~\ref{table:bin}--\ref{table:ambig} may be biased to $\chi_{color}^2 > 20$.

\begin{figure}
\plotone{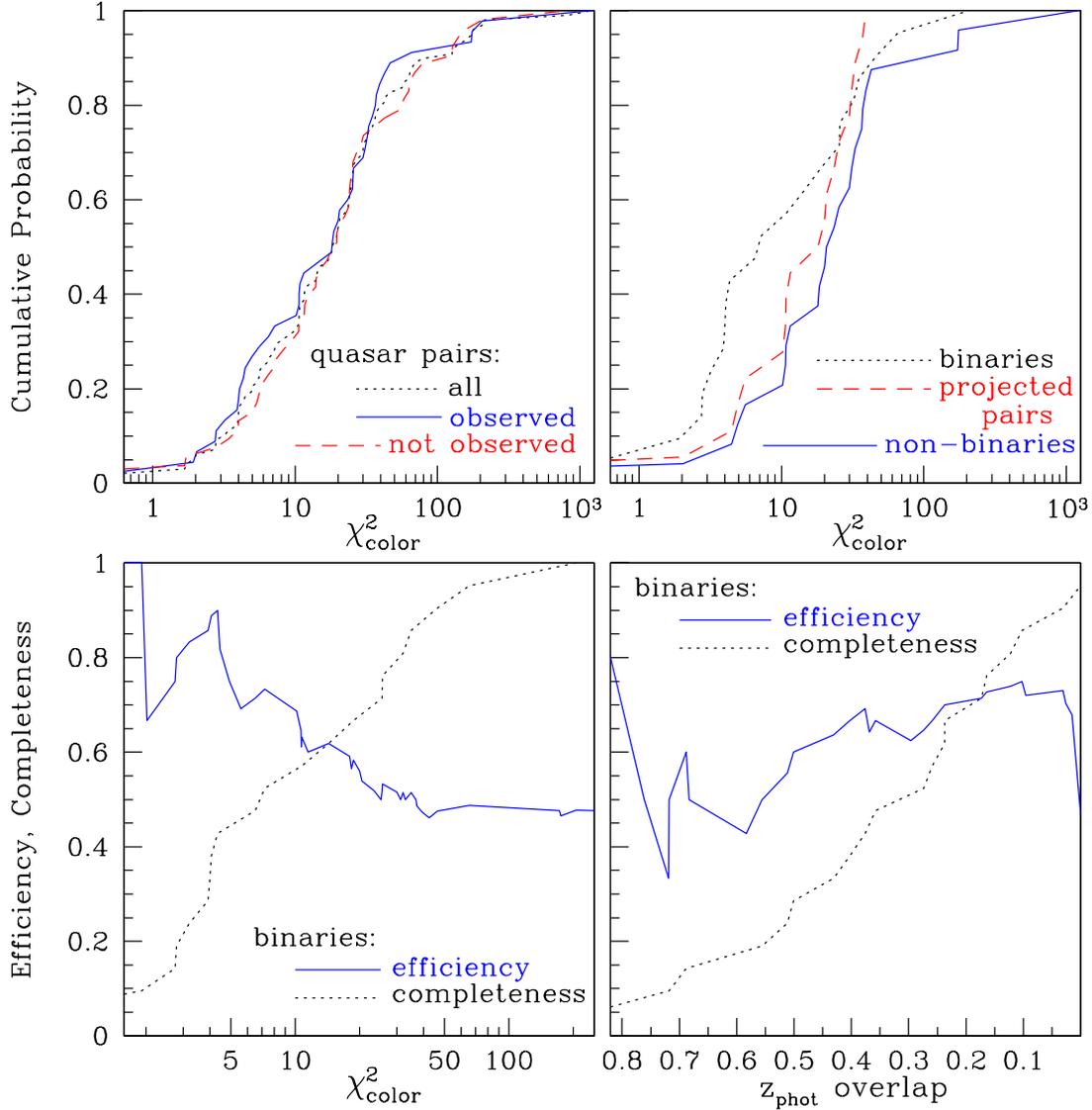}
\caption{{\it Upper-left:} The cumulative probability distribution of the $\chi_{color}^2$ color similarity statistic (Equation~\ref{eqn:chi}) for the 45 spectroscopically observed DR4 KDE candidate binary quasars, the 53 that are not yet observed, and for all 98 candidate pairs. {\it Upper-right:} Similar to upper-left, but for the 21/45 observed candidates that are binary quasars (or lenses), for the 24/45 that are not binaries and for the 18/45 quasar pairs at disjoint redshifts. {\it Lower-left:} The selection completeness and efficiency effects of imposing a $\chi_{color}^2$ limit on the 45 observed DR4 KDE candidate binaries. Efficiency is (number of binaries $< \chi_{color}^2$)/(total candidates $< \chi_{color}^2$); completeness is (number of binaries or lenses $< \chi_{color}^2$)/21. {\it Lower-right:} Similar to lower-left, but using the fractional overlap of the photometric redshift solutions for the candidate components ($z_{phot}$) as the determinant of binary selection. Note that more overlap in photometric redshift implies greater color similarity.
\label{fig:col}}
\end{figure}

In the upper-left panel of Figure~\ref{fig:col}, we demonstrate that the $\chi_{color}^2$ values of the DR4 KDE binary candidates identified to date fairly represent the full sample. We compare the cumulative fraction of the 45 identified candidates (i.e., Table~\ref{table:bkdown}) to the remaining 53 ($3\arcsec < \Delta\theta < 6\arcsec$) candidates, as a function of $\chi_{color}^2$. A two-sample Kolmogorov-Smirnov test cannot distinguish the distributions, suggesting that the colors of the observed candidates fairly represent all candidates. The upper-right panel of Figure~\ref{fig:col} compares the $\chi_{color}^2$ cumulative probability for the 21 confirmed binary quasars (or lenses) and the 24 confirmed non-binaries (projected quasar pairs, star-quasar pairs, NELG-quasar pairs). The K-S test probability that these two distributions are drawn from the same underlying $\chi_{color}^2$ distribution is $\sim10\%$, suggesting that $\chi_{color}^2$ can indeed discriminate binary quasars from non-binaries.

As our candidates are selected without a $\chi_{color}^2$ cut, we can ask what $\chi_{color}^2$ limit optimizes completeness (number of binaries/21 total binaries) and efficiency (number of binaries/45 observed candidates). To sample $\geqsim~50\%$ ($\geqsim~66\%$) of binary quasars requires $\chi_{color}^2~\leqsim~10$ ($\chi_{color}^2~\leqsim~20$), efficient at 70\% (55\%), while $\chi_{color}^2~\leqsim~70$--100, contains 95\% of all binaries (only missing the ``ambiguous'' SDSSJ1235+6836A,B)  and remains 50\% efficient. A $\chi_{color}^2 < 70$ cut rejects all but one quasar-star pair, while retaining all quasar-quasar projections. 

As there is reasonable scatter in the quasar color-redshift relation (e.g., \citealt{Ric01}) full photometric redshift (henceforth {\it photoz}) information should better select binary quasars. To test this, we consider the primary ``CZR'' {\it photoz} solution (\citealt{Wei04}; ``$z_{phot}$ range'' in Tables 1--2) for each quasar in our sample. We determine the overlap fraction of the primary {\it photoz} solutions of the two quasars in a pair, multiplying by the probability that the quasar occupies that primary peak. 

The completeness and efficiency of a binary quasar sample obtained by considering {\it photoz} overlap are plotted in the lower-left panel of Figure~\ref{fig:col}. Confirmed non-binaries typically have no {\it photoz} overlap. A probability cut at $>3\%$ overlap will return 90\% of binaries and is 73\% efficient. The two binaries that are missed are the ``ambiguous'' SDSSJ1235+6836A,B, and SDSSJ1637+2636A,B, the ``A'' component of which has a poorly behaved {\it photoz} solution. If one additionally observed all candidates that contained a quasar with a poor {\it photoz} (characterized by a probability of $< 0.5$), 95\% of binary quasars would be observed at 69\% efficiency. Of course, although cutting on {\it photoz} overlap is a more efficient mechanism for selecting binaries, it does so at the expense of projected quasar pairs. A cut at $<3\%$ overlap could therefore be used to discard binary quasars in favor of projected pairs.

In conclusion, if a survey's goal is to select both binary quasars and projected quasar pairs, a cut of $\chi_{color}^2~\leqsim~70$ (after first applying an efficient photometric quasar classification technique such as KDE) will return 97\% of binaries, lenses and projected pairs (multiplied by, e.g., the 95\% completeness of the KDE technique itself), at 93\% efficiency. This is hardly surprising, and simply further confirms that the KDE technique is $\sim95\%$ efficient \citep{Ric04,Mye06,Mye07a}. Cuts of $\chi_{color}^2~\leqsim~20$, or stricter, may be necessary without prior KDE photometric classification but will miss $\geqsim~33\%$ of all binary quasars. Interestingly, 2 ($\sim10\%$) of the DR4 KDE binary quasars (and one lens!) have more dissimilar colors than any of the quasar-quasar projections, hinting that physical interactions may affect the colors of a few binary quasars. If a survey goal is observing only binary quasars, a cut of $>0.03$ in the overlap of the two quasars' {\it photozs} is superior to $\chi_{color}^2$ selection. We stress that this analysis applies only to quasars pre-selected using an efficient photometric classification technique. 

\section{Projected Quasar Clustering at $25\kpch$}
\label{sec:clus}

With our new observations (Table~\ref{table:bin} and \ref{table:ambig}), we have now identified all DR4 KDE binary quasar candidates with $A_g < 0.17$, $g < 20.85$ and $3.9\arcsec < \Delta\theta < 5.2\arcsec$.  In Table~\ref{table:clus} we compile the binaries that meet these criteria. The three possible lenses in this subsample are wide-separation ($\Delta\theta \geq 4.5\arcsec$), making them likely binaries. As our sample is complete over $3.9\arcsec < \Delta\theta < 5.2\arcsec$, it is also effectively complete for quasars in SDSS DR4 with separations on proper scales of $23.7 < R_{prop} < 29.7\kpch$ over $1.0 < z < 2.1$.

We study quasar clustering using the $DD/DR$ estimator (e.g., \citealt{Sha83}) for quasar-quasar  ($QQ$) pair counts compared to expected quasar-random ($QR$) pair counts

\beq
\overline{W_p} = \frac{QQ}{\langle QR \rangle} - 1
\label{eqn:wp}
\eeq

\noindent Higher-order corrections (e.g., \citealt{Lan93}), reduce to Equation~\ref{eqn:wp} for the small scales and large volumes we consider. We use small-number Poisson errors from \citet{Geh86}. Poisson errors are valid on small scales where pair counts are independent (e.g., \citealt{Cro96,Mye06}).

As our clustering sample is a subset of all SDSS DR4 photometric objects, our selection function is very simple. We calculate $\langle QR \rangle$ in Equation~\ref{eqn:wp} by constructing a catalog of random points with the same angular coverage as SDSS DR4, correcting the SDSS data for mask holes, as in \citet{Mye06,Mye07a}. We further limit our random catalog to areas of the sky with Galactic absorption $A_g < 0.17$. We assign random points a redshift according to a fit to the normalized redshift distribution of ($A_g < 0.17$, $g < 20.85$) quasars in the DR1 catalog \citep{Sch03}, from which the DR4 KDE quasar classification is trained. Figure 7 of \citet{Mye06} is similar to this redshift distribution, and \citet{Mye06} argue that including additional quasars that overlap the KDE color space minimally impacts this $N(z)$ distribution. To represent our fit, we use a modified Gaussian

\beq
{\rm d}N = \beta\exp\frac{-\left|z-\bar{z}\right|^{n}}{n\sigma_i^{n}}{\rm d}z
\label{eqn:Nz}
\eeq

\noindent and find a best fit with $\bar{z}=1.4$, $\sigma=0.6$, $\beta=0.65$ and $n=3$ (see also \citealt{Mye07a}). We have repeated our analyses instead using a spline fit, and our results differ by $\leqsim~2$\%.

To determine $\langle QR \rangle$ in a given bin we use a random catalog with 1000 times as many points as the ($A_g < 0.17$, $g < 20.85$) KDE DR4 photometric quasar catalog. We total all $QR$ counts in the \textit{angular} bin of interest, and normalize the result (divide by 1000). We then create a random catalog with points distributed according to Equation~\ref{eqn:Nz}, and determine the fraction of pairs that would lie within $|\Delta v_{\|}|<2000\kms$ by Monte Carlo sampling to 0.1\% precision. Multiplying this fraction by the normalized $QR$ counts yields $\langle QR \rangle$. In a bin of $3.9\arcsec < \Delta\theta < 5.2\arcsec$ we expect 30.3 $QR$ counts. Over our full $N(z)$, we expect 0.0166 pairs with $|\Delta v_{\|}|<2000\kms$. Thus $\langle QR \rangle=0.503$, compared to $QQ=16$ for the 8 (non-unique) pairs in Table~\ref{table:clus}. The implied projected correlation function averaged over $3.9\arcsec < \Delta\theta < 5.2\arcsec$ is thus $\overline{W_p}=30.8 \pm^{15.7}_{11.0}$. We note that, for all candidate pairs, $QQ(3.9\arcsec < \theta < 5.2\arcsec)=40$, yielding an angular correlation of $\omega(\theta) = (40/30.3)-1= 0.320\pm^{0.366}_{0.292}$, consistent with M07b. 

One of the pairs (SDSSJ1507+2903) in Table~\ref{table:clus} has $|\Delta v_{\|}|<2100\kms$. We reasonably include this pair in our analysis, given the precision of our redshift estimates. Instead rejecting SDSSJ1507+2903 and assuming $QQ=14$ would lower our estimate of $\overline{W_p}$ by $\sim10\%$. The $|\Delta v_{\|}|<2000\kms$ limit from H06 is intended to bracket possible shifts in quasar lines, and thus incorporate possible errors on $|\Delta v_{\|}|$ but is otherwise arbitrary. Although our Monte Carlo sampling of $|\Delta v_{\|}|<2000\kms$ pairs does not model any error in $|\Delta v_{\|}|$, relaxing this velocity window to $|\Delta v_{\|}|<2100\kms$ would imply $\langle QR \rangle=0.533$ for $QQ=16$, lowering our $\overline{W_p}$ estimate by $\sim5\%$. As the change in $\overline{W_p}$ implied by relaxing our velocity criterion is far smaller than our errors on $\overline{W_p}$, we proceed including SDSSJ1507+2903 in our analyses and maintaining $|\Delta v_{\|}|<2000\kms$ in our Monte Carlo sampling.

\begin{figure}
\plottwo{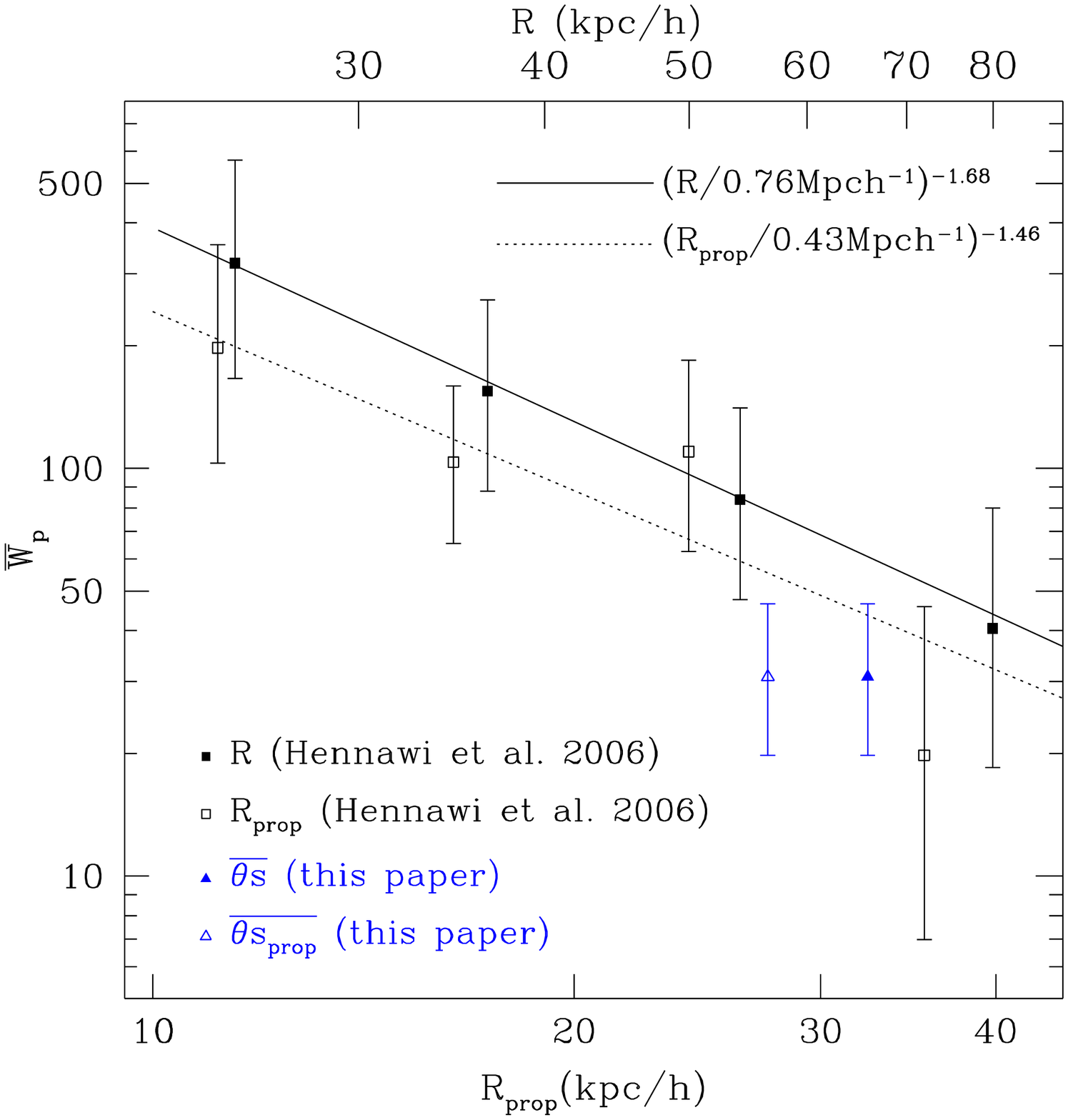}{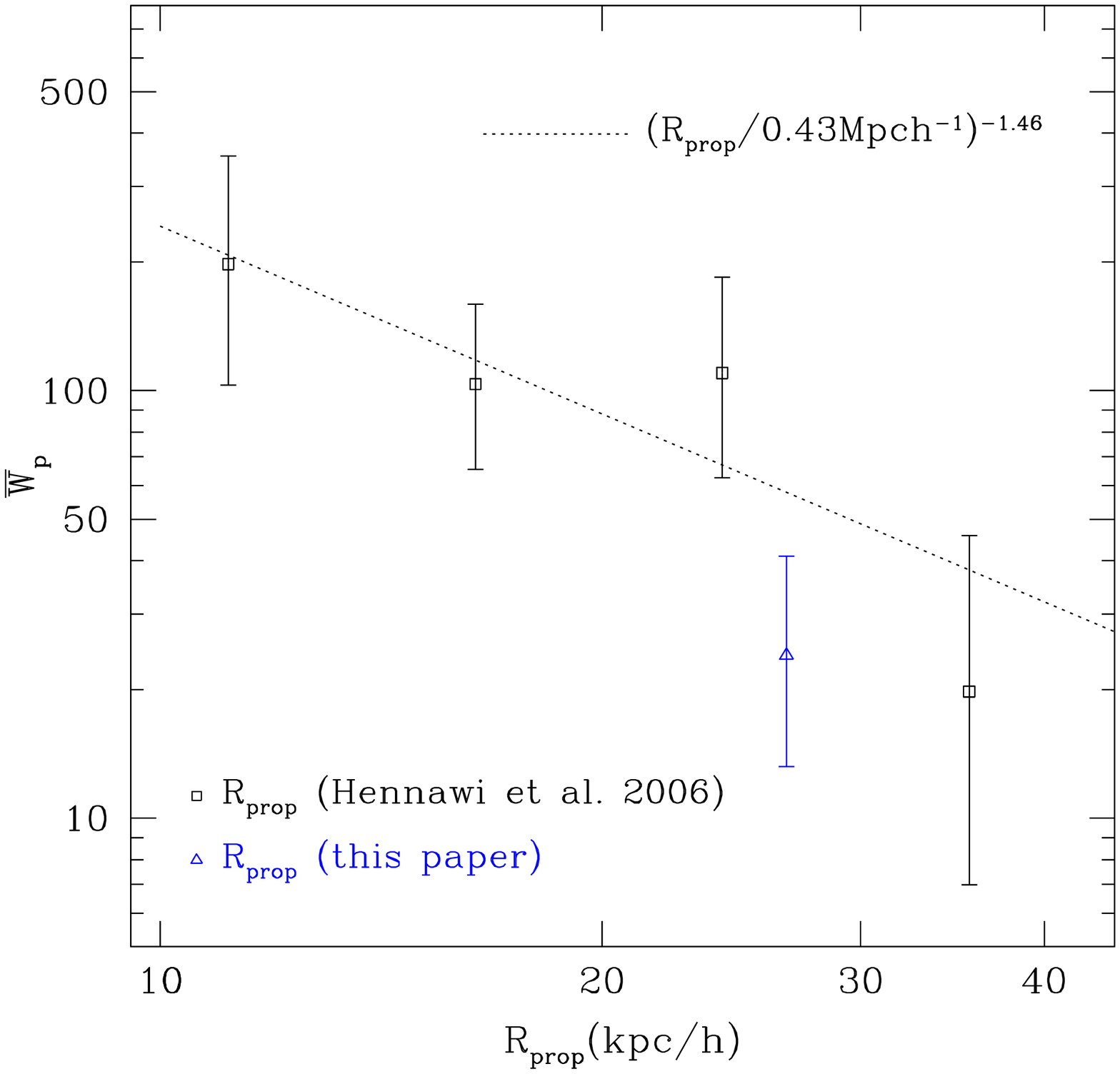}
\caption{The projected correlation function, $\overline{W_p}$, of quasars on small scales.  {\it Left:} The results of H06 in proper (lower axis) and comoving (upper axis) coordinates are compared to our mean DR4 KDE spectroscopic results averaged over $3.9\arcsec < \Delta\theta < 5.2\arcsec$ and projected to a mean transverse separation at $z=1.4$ (the mean redshift of both the DR4 KDE quasar sample and the subset of 16 quasars listed in Table~\ref{table:clus}), in proper (open triangle) and comoving (solid triangle) coordinates. The dashed (solid) line shows a power law fit to the H06 data on proper (comoving) scales $<100\kpch$ ($<200\kpch$). {\it Right:} A similar comparison but our data are now averaged over scales of $23.7 < R_{prop} < 29.7\kpch$ and redshifts of $1.0 < z < 2.1$. The DR4 KDE binary quasar sample is spectroscopically complete for $A_g < 0.17$, $g < 20.85$, and $3.9\arcsec < \Delta\theta < 5.2\arcsec$, and for $23.7 < R_{prop} < 29.7\kpch$ over $1.0 < z < 2.1$.
\label{fig:clus}}
\end{figure}

We apply our approach to different redshift ranges by weighting the number density of points in the initial calculation of $QR$ by the relative fraction of DR4 KDE quasars obtained by integrating under Equation~\ref{eqn:Nz}. We can thus determine $\langle QR \rangle$ over $1.0 < z < 2.1$, for which our sample is spatially complete on scales of $23.7 < R_{prop} < 29.7\kpch$. In Figure~\ref{fig:clus} we compare the small-scale clustering of quasars determined from our complete DR4 binary quasar sample to the results from H06. In the left-hand panel, we consider our calculation for all quasars in Table~\ref{table:clus}, and project the result for both proper and comoving scales by placing all binaries at the mean proper or comoving distance of our sample. In the right-hand panel, we consider binaries in Table~\ref{table:clus} with $23.7 < R_{prop} < 29.7\kpch$ and $1.0 < z < 2.1$. We additionally include SDSSJ1635+2911 in this second sample, as we are spatially complete for component separations of $23.7 < R_{prop} < 29.9$ for  $1.03 < z < 2.1$.

Figure~\ref{fig:clus} demonstrates that our results, using a complete, statistical sample of UVX binary quasars are broadly consistent with H06. To compare with H06 we fit power laws, displayed in Figure~\ref{fig:clus}, out to proper (comoving) scales of $<100\kpch$ ($<200\kpch$), and integrate them over our scales of interest. At the mean redshift ($\bar{z}=1.40$) of our sample, $3.9\arcsec < \theta < 5.2\arcsec$, the range for which our sample is complete, is equivalent to proper (comoving) scales of $23.5\kpch < R_{prop} < 31.4\kpch$ ($56.5\kpch < R < 75.3\kpch$).  Over this angular range, we find $\overline{W_p}=30.8 \pm^{15.7}_{11.0}$ for our data. The proper (comoving) power-law fit to the data from H06, implies $\overline{W_p} = 55.1$ (60.4) over the same scales, a $1.5\sigma$ ($1.9\sigma$) difference. The difference is slightly more pronounced if we determine $\overline{W_p}$ for the H06 data at the mean scale of the 8 binaries in our sample, instead of projecting $3.9\arcsec < \theta < 5.2\arcsec$ back to $z=1.4$. For the 5 binaries in our spatially complete clustering subsample, which covers scales of $23.7\kpch < R_{prop} < 29.9\kpch$, we find $\overline{W_p}=24.0 \pm^{16.9}_{10.8}$, and the H06 data implies $\overline{W_p} = 57.2$, 2.4~times (and $2.0\sigma$) higher than our result.

\section{Discussion}

We find that the projected correlation function of quasars at proper scales of $\sim25\kpch$ has an amplitude a factor of 2.4~times lower than that determined by H06. H06 argue that clustering on these scales is $\sim10$ times higher than expected from projecting the quasar autocorrelation of \citet{Por04} to smaller scales. Our data thus imply excess quasar clustering at $\sim25\kpch$ of a factor of $\sim4$, consistent with the quoted upper limit in M07b. Given that our sample is targeted differently than any previous samples of binary quasars (i.e. UVX but otherwise regardless of the color similarity of the candidate components), and given our simple selection function, our work might be viewed as independently corroborating the evidence for excess quasar clustering on small scales first detected in H06. We note that our binary quasar clustering subsample is largely independent of the sample used by H06, as of the 8 binaries listed in Table~\ref{table:clus} only 2 appear in Table~\ref{table:old}.

At $R_{prop}\sim25\kpch$ we find a significantly smaller clustering amplitude than found by H06. As the binaries in Table~\ref{table:clus} are all at $z < 2$ but H06 include several $R_{prop}\sim25\kpch$ binaries with $z > 2$, an interesting possibility is that $\overline{W_p}$ is a function of scale {\it and} redshift. Certainly, on large scales ($\geqsim~1\Mpch$), quasars cluster twice as strongly at $z > 2$ than at $z < 1$ (e.g., \citealt{Cro05,She07}). The mean redshift of the H06 sample is $\bar{z} = 1.87$ (J. Hennawi 2007, private communication) compared to $\bar{z} = 1.42$ for our sample in Table~\ref{table:clus}. \citet{Cro05} estimate quasar bias on large (i.e., $\geqsim~1\Mpch$) scales to follow $b(z) = 0.53 + 0.289(1+ z)^2$. Given that quasar clustering scales as $b^2$, the implied relative amplitude between quasar clustering in H06 and our sample is $\sim1.72$. Scaling our measurement of $\overline{W_p}=30.8 \pm^{15.7}_{11.0}$ by this factor implies $\overline{W_p}=53.0 \pm^{27.0}_{18.9}$, easily consistent with the value of $\overline{W_p} = 57.2$ implied by our fit to H06 in Figure~\ref{fig:clus}. Scaling the $\overline{W_p}=24.0 \pm^{16.9}_{10.8}$ estimate for our complete subsample ($\bar{z}=1.60$) in the same way implies $\overline{W_p}=33.1 \pm^{23.4}_{14.9}$, lower than $\overline{W_p} = 57.2$ but consistent to $\sim1\sigma$. Thus, multiplying our quasar clustering amplitudes on {\it small} scales by the implied bias evolution from quasar clustering on {\it large} scales somewhat reconciles our clustering amplitudes with the higher redshift results from H06. Consistent evolution of quasar clustering on large and small scales is a natural feature of merger-driven models of quasar origin (e.g., Figure~17 of \citealt{Hop07pre}).

We can also look for direct redshift dependence within our sample. Splitting the sample in Table~\ref{table:clus} at $z=1.55$, we find $\overline{W_p} = 27.9\pm^{22.9}_{13.9}$ for $z < 1.55$ ($\bar{z}\sim1.07$) and $\overline{W_p} = 34.9\pm^{28.4}_{17.2}$ for $z > 1.55$ ($\bar{z}\sim1.76$). Although these results seem consistent, the mean separation of the binaries at $z < 1.55$ ($z > 1.55$) is $R_{prop} = 23.8$ (28.4). The power law fits displayed in Figure~\ref{fig:clus} imply that we should therefore expect quasars to be $\sim30$\% \textit{more} clustered at $z < 1.55$ than at $z > 1.55$. Incorporating this $\sim30$\% correction, we measure the ratio of the amplitudes of $\overline{W_p}$ at $z > 1.55$ and $z < 1.55$ to be $1.62\pm1.54$. Although not a very significant detection, a factor of $1.62\pm1.54$ is again consistent with quasar clustering evolution on large scales, as a $b(z) = 0.53 + 0.289(1+ z)^2$ bias relation implies a ratio of $\sim2.4$ between the amplitudes of $\overline{W_p}$ at $z = 1.76$ and $z = 1.07$.

Our analyses could be easily extended, as the 98 ($3\arcsec\leq\Delta\theta<6\arcsec$) DR4 KDE binary quasar candidates should contain $\sim40$ binary quasars.  Based on our $\overline{W_p}$ estimates, such a sample is the minimum necessary to detect any redshift dependence to quasar clustering at $R_{prop}\sim25\kpch$, providing only a $\sim1.5\sigma$ detection. Completing this sample on a 4-meter class telescope would likely require dark time, as 14 of the DR4 binary candidates have a component at $g > 21$. Alternatively, \citet{She07} suggest a survey of quasar pairs at $z > 3$, where stronger quasar clustering may lead to a more significant detection of evolution. The DR6 KDE quasar catalog (Richards et al, in prep), which will be selected to $i < 21$ ($g~\leqsim~21.25$), should contain $\sim500$ binary quasar candidates across a large redshift range ($0.4~\leqsim~z~\leqsim~5$) with which to pursue this goal in a single homogeneous sample.

\section{Conclusions}

We have presented a spectroscopic survey of binary quasar candidates with separations in the range $3\arcsec\leq\Delta\theta<6\arcsec$. Our candidates (see Table~1 of M07b) are a subsample of quasars photometrically classified in SDSS DR4 using the KDE technique of \citet{Ric04}. We define a binary as a quasar pair with a line-of-sight velocity difference of $|\Delta v_{\|}|<2000\kms$ (see H06). We present 27 new sets of observations and identify both members of 22 candidate binary quasars. Of the 22 new pairs, 10 turn out to be binary quasars (of which $\sim2$ might actually be lenses). This roughly doubles the number of known binary quasars with $3\arcsec\leq\Delta\theta<6\arcsec$ at $z~\leqsim~2$ and $g < 21$. A further 9 of our observed candidates are projected quasar pairs, and 3 contain a NELG or star. This confirms that the KDE technique is $\sim95$\% efficient at selecting quasars (e.g., \citealt{Ric04,Mye06,Mye07a}). Combined with observations from the literature (mainly from H06), 46\% of the DR4 KDE binary quasar candidates have now been observed, of which $\sim47\%$ are binaries or lenses, $\sim40\%$ are projected quasar pairs, and the remainder contain a non-quasar.

As our candidate binaries, beyond a UVX cut, are selected regardless of the {\it relative} colors of the quasars in the pair, we can try to assess the color similarity criteria that optimally select binary quasars. For quasars pre-selected with an already efficient approach such as the KDE technique, we find that a $\chi^2$ color similarity statistic of $\chi_{color}^2 < 70$ will return 97\% of binaries, lenses and projected pairs (multiplied by the 95\% completeness of the KDE technique itself) at 93\% efficiency. Most of this efficiency in selecting quasar pairs comes from the KDE technique itself, as imposing no color similarity criterion is $87\%$ efficient. To select binary quasars, while rejecting projected quasar pairs, we recommend a cut in the overlap of the photometric redshifts of the two candidate quasars in a pair. An overlap of $\geqsim~0.03$ in the primary solution for the photometric redshift probability density functions of the pair can be constructed to be $\sim95\%$ complete and $\sim70\%$ efficient for binary quasars. Similarly, of course, the reverse probability cut of $\leqsim~0.03$, perhaps coupled with a $\chi_{color}^2 < 70$ cut to remove stars, can be used to reject binary quasars in favor of projected quasar pairs.

We measure the clustering of a complete sample of DR4 binaries on proper scales of $23.7 < R_{prop} < 29.7\kpch$. We find that, at $\sim25\kpch$, quasars cluster with an amplitude 2.4~times, or $2.0\sigma$, lower than determined by H06. As the mean redshift of the H06 sample is $\bar{z} = 1.87$ compared to $\bar{z} = 1.40$ for our sample, this can be interpreted as evidence of evolution in quasar clustering on scales of $\sim25\kpch$. The implied evolution is broadly consistent with merger-driven models, where the quasar population is expected to evolve with consistent large-to-small scale clustering (e.g., \citealt{Hop07pre}). We find no significant evidence for quasar clustering evolution at $\sim25\kpch$, from $\bar{z}\sim1.07$ to $\bar{z}\sim1.76$, in our sample alone. Assuming evolution in the binary quasar population at the level suggested by our current sample, we argue that the 98 quasars in the DR4 KDE candidate binary sample will detect any clustering evolution at proper scales of $\sim25\kpch$ at $\sim1.5\sigma$ significance. A sample of $\sim200$ candidates ($\sim80$ binary quasars) will be necessary to definitively detect clustering evolution at $\sim25\kpch$ for $z~\leqsim~2.5$.

\acknowledgments

We thank the NOAO staff for their indispensable help and knowledge. In particular, we thank Buell Januzzi, without whose support this work would not have been possible in a timely fashion. ADM and RJB acknowledge support from NASA through grant NN6066H156, from Microsoft Research, and from the University of Illinois. DPS acknowledges NSF support through grant AST-0607634. PBH is supported by NSERC. The authors made extensive use of the storage and computing facilities at the National Center for Supercomputing Applications and thank the technical staff for their assistance in enabling this work. We thank Robert Nichol and Alex Gray for their invaluable work on the KDE catalog.

Funding for the SDSS and SDSS-II has been provided by the Alfred P. Sloan Foundation, the Participating Institutions, the National Science Foundation, the U.S. Department of Energy, the National Aeronautics and Space Administration, the Japanese Monbukagakusho, the Max Planck Society, and the Higher Education Funding Council for England. The SDSS Web Site is http://www.sdss.org/.

The SDSS is managed by the Astrophysical Research Consortium for the Participating Institutions. The Participating Institutions are the American Museum of Natural History, Astrophysical Institute Potsdam, University of Basel, Cambridge University, Case Western Reserve University, University of Chicago, Drexel University, Fermilab, the Institute for Advanced Study, the Japan Participation Group, Johns Hopkins University, the Joint Institute for Nuclear Astrophysics, the Kavli Institute for Particle Astrophysics and Cosmology, the Korean Scientist Group, the Chinese Academy of Sciences (LAMOST), Los Alamos National Laboratory, the Max-Planck-Institute for Astronomy (MPIA), the Max-Planck-Institute for Astrophysics (MPA), New Mexico State University, Ohio State University, University of Pittsburgh, University of Portsmouth, Princeton University, the United States Naval Observatory, and the University of Washington.

\clearpage

\renewcommand{\thetable}{1\alph{table}}

\begin{deluxetable}{crcccccll}
\tabletypesize{\scriptsize}
\tablecaption{DR4 KDE candidate binaries for which we have observed one member\label{table:singles}}
\tablecolumns{9}
\tablewidth{0pt}
\tablehead{
\colhead{$\Delta\theta$} & \colhead{$\chi_{color}^2$} & \colhead{Name} & \colhead{$\alpha$ (J2000)} & \colhead{$\delta$ (J2000)} &\colhead{$g$}  & \colhead{$z_{phot}$ range} & \colhead{SDSS z} & \colhead{Our z}
}
\startdata
5.50 &   36.2 & SDSSJ093014.81+420038.7 & 09 30 14.816 & +42 00 38.71 & 19.71 & 1.15,1.425,1.50,0.93 &       & 0.544\tablenotemark{?} \\   
     &        & SDSSJ093015.01+420033.6 & 09 30 15.016 & +42 00 33.68 & 20.03 & 1.85,2.025,2.15,0.62 &       &      \\ 
\hline
5.55 &   24.2 & SDSSJ093521.02+641219.8 & 09 35 21.020 & +64 12 19.89 & 20.99 & 1.45,1.775,2.10,0.92 &       & 1.566  \\   
     &        & SDSSJ093521.80+641221.9 & 09 35 21.807 & +64 12 22.00 & 20.96 & 0.25,0.375,0.45,0.66 &       &      \\ 
\hline
5.33 &    6.4 & SDSSJ095840.74+332216.3 & 09 58 40.746 & +33 22 16.31 & 19.18 & 1.35,1.475,2.10,0.84 & 1.891 & 1.888  \\   
     &        & SDSSJ095840.94+332211.5 & 09 58 40.945 & +33 22 11.59 & 20.64 & 1.45,1.725,2.10,0.91 &       &      \\ 
\hline
3.46 &  161 & SDSSJ103939.31+100253.0 & 10 39 39.317 & +10 02 53.01 & 18.42 & 0.10,0.175,0.25,0.56 & 0.161 & 0.161  \\   
     &        & SDSSJ103939.53+100254.3 & 10 39 39.532 & +10 02 54.40 & 19.60 & 0.65,1.125,1.55,0.92 &       &      \\ 
\hline
4.22 &    8.2 & SDSSJ162847.75+413045.4 & 16 28 47.752 & +41 30 45.45 & 19.81 & 1.35,1.525,1.70,0.85 &       &      \\   
     &        & SDSSJ162848.06+413043.1 & 16 28 48.069 & +41 30 43.19 & 20.40 & 1.95,2.075,2.20,0.63 &       & 0.831\tablenotemark{?} \\ 
\hline
\enddata

\tablecomments{The angular pair separations are denoted $\Delta\theta$ ($\arcsec$), and $\chi_{color}^2$ is each pair's color similarity statistic (Equation~\ref{eqn:chi}). The photometric redshifts $z_{phot}$ are expressed as ``lowest extent, peak, highest extent, probability of true redshift lying in this range". The ``SDSS $z$'' column shows matches to any spectroscopic object in the SDSS DR6 Catalog Archive Server (mainly, e.g., \citealt{Sch07}). The ``Our z'' column lists our new spectroscopic confirmations from KPNO data. In the ``SDSS z'' and ``Our z'' columns the object is a quasar at the provided redshift, unless otherwise noted. Redshifts labeled with a ? are based on a single emission line, which is reasonably assumed to be \MgII. $g$ is not corrected for Galactic extinction.
}
\end{deluxetable}

\begin{deluxetable}{crcccccll}
\tabletypesize{\scriptsize}
\tablecaption{Confirmed binary quasars in the DR4 KDE candidate sample \label{table:bin}}
\tablecolumns{9}
\tablewidth{0pt}
\tablehead{
\colhead{$\Delta\theta$} & \colhead{$\chi_{color}^2$} &
\colhead{Name} & \colhead{$\alpha$ (J2000)} & \colhead{$\delta$
(J2000)} & \colhead{$g$}  & \colhead{$z_{phot}$ range} & \colhead{SDSS z} & \colhead{Our z} 
}
\startdata
*3.56 &    4.0& SDSSJ1158+1235A & 11 58 22.776 & +12 35 18.59 & 19.90 & 0.45,0.525,0.65,0.55 &      & 0.596\tablenotemark{?} \\   
      &       & SDSSJ1158+1235B & 11 58 22.989 & +12 35 20.31 & 20.12 & 0.40,0.475,0.65,0.63 &       & 0.596\tablenotemark{?} \\ 
\hline
*4.74 &   32.0& SDSSJ1320+3056A & 13 20 22.545 & +30 56 22.87 & 18.60 & 1.35,1.475,1.65,0.90 &  1.597    & 1.595  \\   
      &       & SDSSJ1320+3056B & 13 20 22.643 & +30 56 18.29 & 19.92 & 1.45,1.575,1.80,0.51 &       & 1.596  \\ 
\hline
*4.50 &    2.8 & SDSSJ1418+2441A & 14 18 55.418 & +24 41 08.92 & 19.27 & 0.45,0.525,0.70,0.96 &  0.573     & 0.572  \\   
     &        & SDSSJ1418+2441B & 14 18 55.536 & +24 41 04.71 & 20.22 & 0.40,0.625,0.70,0.86 &       & 0.573  \\ 
\hline
4.27 &    3.2 & SDSSJ1426+0719A & 14 26 04.266 & +07 19 25.86 & 20.82 & 0.95,1.175,1.45,0.99 &       & 1.312  \\   
     &        & SDSSJ1426+0719B & 14 26 04.326 & +07 19 30.04 & 20.12 & 1.00,1.225,1.45,0.97 &       & 1.309  \\ 
\hline
5.41 &   25.6 & SDSSJ1430+0714A & 14 30 02.664 & +07 14 15.62 & 20.27 & 1.00,1.225,1.40,0.97 &       & 1.246  \\   
     &        & SDSSJ1430+0714B & 14 30 02.886 & +07 14 11.33 & 19.50 & 1.05,1.375,1.45,0.97 &  1.258     & 1.261  \\ 
\hline
5.14 &   65.6 & SDSSJ1458+5448A & 14 58 26.165 & +54 48 14.85 & 20.79 & 1.50,1.775,1.95,0.75 &       & 1.913  \\   
     &        & SDSSJ1458+5448B & 14 58 26.728 & +54 48 13.19 & 20.53 & 1.65,1.925,1.95,0.47 &       & 1.912 \\ 
\hline
*3.45 &  10.8 & SDSSJ1606+2900A & 16 06 02.812 & +29 00 48.79 & 18.50 & 0.50,0.725,1.00,0.65 & 0.770  & 0.769?  \\   
      &       & SDSSJ1606+2900B & 16 06 03.021 & +29 00 50.88 & 18.42 & 0.70,0.875,1.00,0.92 &       & 0.769?  \\ 
\hline
*4.92 &   25.6& SDSSJ1635+2911A & 16 35 10.148 & +29 11 20.65 & 18.83 & 1.45,1.575,1.80,0.84 & 1.586 & 1.582 \\   
      &       & SDSSfJ1635+2911B & 16 35 10.306 & +29 11 16.19 & 20.43 & 1.40,1.525,1.85,0.79 &       & 1.590  \\ 
\hline
\enddata

\tablecomments{
We define a binary quasar by a line-of-sight velocity difference of $|\Delta v_{\|}|<2000\kms$ in the rest-frame of either component. The pair containing SDSSJ143002.66+071415.6 has velocity difference $|\Delta v_{\|}|=2000\pm400\kms$, just inside our definition of a binary. Components are denoted A and B so that the position angle from A to B lies between $0^\circ$ and $180^\circ$. A preceding * denotes that our spectroscopy alone is insufficient to rule out a lens interpretation for this pair (see \S\ref{sec:Lens}). SDSSJ1320+3056A first appeared with a confirmed redshift (z=1.587) in \citet{Ver04}. $g$ is not corrected for Galactic extinction. {\em See Table~\ref{table:singles} for additional notes describing shared notation}.}

\end{deluxetable}


\begin{deluxetable}{crcccccll}
\tabletypesize{\scriptsize}
\tablecaption{Confirmed projected pairs in the DR4 KDE candidate sample \label{table:nonbin}}
\tablecolumns{9}
\tablewidth{0pt}
\tablehead{
\colhead{$\Delta\theta$} & \colhead{$\chi_{color}^2$} & \colhead{Name} & \colhead{$\alpha$ (J2000)} & \colhead{$\delta$ (J2000)} &\colhead{$g$}  & \colhead{$z_{phot}$ range} & \colhead{SDSS z} & \colhead{Our z}
}
\startdata
4.28 & 1220 & SDSSJ083258.34+323003.3 & 08 32 58.348 & +32 30 03.35 & 19.96 & 2.75,2.775,2.80,0.88 &           & 0.397  \\   
     &        & SDSSJ083258.56+323000.0 & 08 32 58.567 & +32 30 00.08 & 19.59 & 0.40,0.425,0.50,0.70 &           & star   \\ 
\hline
3.42 &    2.0 & SDSSJ084257.37+473342.5 & 08 42 57.378 & +47 33 42.56 & 19.00 & 0.50,0.625,0.70,0.51 & 1.552 & 1.554  \\   
     &        & SDSSJ084257.63+473344.7 & 08 42 57.638 & +47 33 44.74 & 20.45 & 0.85,1.775,2.10,0.84 &       & 1.681  \\ 
\hline
4.28 &    4.9 & SDSSJ085914.77+424123.6 & 08 59 14.771 & +42 41 23.67 & 21.02 & 0.95,1.375,1.65,0.66 &       & 1.396  \\   
     &        & SDSSJ085915.15+424123.5 & 08 59 15.159 & +42 41 23.58 & 19.22 & 0.80,0.975,1.40,0.93 & 0.898& 0.902\tablenotemark{?}    \\ 
\hline
4.57 &   20.6 & SDSSJ094309.36+103401.3 & 09 43 09.363 & +10 34 01.31 & 20.07 & 0.95,1.525,1.70,0.80 &       & 1.431  \\   
     &        & SDSSJ094309.66+103400.6 & 09 43 09.670 & +10 34 00.65 & 19.24 & 1.05,1.325,1.40,0.99 & 1.238 & 1.240  \\ 
\hline
5.91 &   32.8 & SDSSJ111053.63+605347.9 & 11 10 53.633 & +60 53 47.97 & 18.94 & 0.65,0.775,0.90,0.96 &       & 0.793  \\   
     &        & SDSSJ111054.10+605343.1 & 11 10 54.105 & +60 53 43.16 & 20.86 & 1.55,1.775,2.20,0.68 &       & 0.552\tablenotemark{?} \\ 
\hline
5.19 &   18.4 & SDSSJ112556.32+143148.0 & 11 25 56.321 & +14 31 48.10 & 20.70 & 0.90,1.325,1.60,0.69 &       & NELG \\
     &        & SDSSJ112556.54+143152.1 & 11 25 56.549 & +14 31 52.10 & 20.38 & 1.50,1.725,2.15,0.97 &       & 1.924      \\ 
\hline
5.78 &   29.9 & SDSSJ113637.52+563500.4 & 11 36 37.526 & +56 35 00.48 & 19.94 & 1.85,2.075,2.15,0.63 &       & 1.282  \\   
     &        & SDSSJ113638.09+563503.9 & 11 36 38.090 & +56 35 03.90 & 19.02 & 0.60,0.675,0.90,0.56 &       & 2.672  \\ 
\hline
4.92 &   39.3 & SDSSJ114503.06+660211.3 & 11 45 03.063 & +66 02 11.35 & 20.09 & 1.65,1.825,2.00,0.89 &       & 1.732  \\   
     &        & SDSSJ114503.74+660208.6 & 11 45 03.741 & +66 02 08.67 & 20.24 & 0.50,0.675,0.85,0.52 &       & 2.304  \\ 
\hline
3.26 &   11.5 & SDSSJ123122.27+493433.8 & 12 31 22.276 & +49 34 33.86 & 20.63 & 0.50,0.725,0.90,0.53 &       & 0.780\tablenotemark{?} \\   
     &        & SDSSJ123122.37+493430.7 & 12 31 22.378 & +49 34 30.75 & 19.94 & 1.55,1.775,2.15,0.95 &       & 1.811\tablenotemark{?} \\ 
\hline
4.00 &   10.7 & SDSSJ150656.86+505610.5 & 15 06 56.866 & +50 56 10.56 & 19.22 & 0.65,0.775,1.00,0.75 &       & 0.775\tablenotemark{?} \\   
     &        & SDSSJ150657.18+505607.9 & 15 06 57.183 & +50 56 07.92 & 19.75 & 2.00,2.225,2.40,0.45 &       & 2.204  \\ 
\hline
\enddata

\tablecomments{
SDSSJ112556.32+143148.0, the NELG, has redshift z=0.246. Subsequent to our observations, SDSSJ094309.36+103401.3 appeared in \citet{Ina07}, with $z=1.433$. $g$ is not corrected for Galactic extinction. The redshift for SDSSJ123122.37+493430.7 is based on a single \CIV~emission line, with weak confirming \CIII. {\em See Table~\ref{table:singles} for additional notes describing shared notation}.
}

\end{deluxetable}

\begin{deluxetable}{crcccccll}
\tabletypesize{\scriptsize}
\tablecaption{Ambiguous pairs in the DR4 KDE binary quasar candidate sample \label{table:ambig}}
\tablecolumns{9}
\tablewidth{0pt}
\tablehead{
\colhead{$\Delta\theta$} & \colhead{$\chi_{color}^2$} &
\colhead{Name} & \colhead{$\alpha$ (J2000)} & \colhead{$\delta$
(J2000)} & \colhead{$g$}  & \colhead{$z_{phot}$ range} & \colhead{SDSS z} & \colhead{Our z} 
}

\startdata
4.80 &   42.5 & SDSSJ093424.11+421135.0 & 09 34 24.112 & +42 11 35.05 & 21.01 & 1.45,1.675,2.20,0.90 &       & f/less \\   
     &        & SDSSJ093424.32+421130.8 & 09 34 24.324 & +42 11 30.87 & 20.30 & 1.00,1.125,1.40,0.99 &       & 1.339  \\ 
\hline
3.95 &   10.7 & SDSSJ120727.09+140817.1 & 12 07 27.100 & +14 08 17.18 & 20.39 & 1.60,1.775,2.00,0.89 &       & 1.801\tablenotemark{?}  \\
     &        & SDSSJ120727.25+140820.3 & 12 07 27.259 & +14 08 20.38 & 20.27 & 1.55,1.775,1.95,0.86 &       & 1.8\tablenotemark{??}  \\ 
\hline
3.51 &  207 & SDSSJ1235+6836A & 12 35 54.783 & +68 36 24.78 & 19.04 & 2.75,2.775,2.80,0.55 &       & 1.529\tablenotemark{??} \\   
     &        & SDSSJ1235+6836B & 12 35 55.270 & +68 36 27.07 & 19.70 & 0.50,0.625,1.10,0.51 &       & 1.514  \\ 
\hline
4.35 &   14.4 & SDSSJ1507+2903A & 15 07 46.909 & +29 03 34.15 & 20.44 & 0.80,0.975,1.25,0.77 &       & 0.875\tablenotemark{?} \\   
     &        & SDSSJ1507+2903B & 15 07 47.234 & +29 03 33.28 & 19.97 & 0.70,0.775,0.95,0.66 &       & 0.862\tablenotemark{?} \\ 
\hline
\enddata

\tablecomments{Redshifts marked $??$ are derived from a single emission line. This differs from the $?$ notation as the redshift is based on similar emission in the other component (rather than simply assuming that the emission is \MgII). The ambiguities, and why we conclude that SDSSJ1235+6836 and SDSSJ1507+2903 are binaries but the other pairs are not, are discussed in \S\ref{sec:IntObj}. $g$ is not corrected for Galactic extinction. {\em See Table~\ref{table:singles} for additional notes describing shared notation}.}

\end{deluxetable}


\setcounter{table}{1}
\renewcommand{\thetable}{\arabic{table}}

\begin{deluxetable}{crcccccl}
\tabletypesize{\scriptsize}
\tablecaption{Previously identified DR4 KDE binary quasar candidates ($3\arcsec\leq\Delta\theta<6\arcsec$)\label{table:old}}
\tablecolumns{8}
\tablewidth{0pt}
\tablehead{
\colhead{$\Delta\theta$} & \colhead{$\chi_{color}^2$} &
\colhead{Name} & \colhead{$\alpha$ (J2000)} & \colhead{$\delta$ (J2000)} & \colhead{$g$}  & \colhead{$z_{phot}$ range} & \colhead{z} 
}
\startdata

\sidehead{\underline{Projected pairs}}
4.90 & 36.7 & SDSSJ024907.77+003917.1 & 02 49 07.778 & +00 39 17.12 & 19.36 & 2.00,2.175,2.25,0.48 & 2.164 \\
     &      & SDSSJ024907.86+003912.4 & 02 49 07.866 & +00 39 12.40 & 20.63 & 0.45,0.675,0.85,0.84 & star  \\ \hline
4.09 & 18.0 & SDSSJ083649.45+484150.0 & 08 36 49.456 & +48 41 50.08 & 19.31 & 0.45,0.675,0.80,0.67 & 0.657 \\ 
     &      & SDSSJ083649.55+484154.0 & 08 36 49.554 & +48 41 54.06 & 18.50 & 1.50,1.675,1.95,0.94 & 1.712 \\ \hline
5.42 &  4.5 & SDSSJ090235.35+563751.8 & 09 02 35.356 & +56 37 51.84 & 20.95 & 1.15,1.275,1.45,0.98 & 1.39  \\
     &      & SDSSJ090235.73+563756.2 & 09 02 35.731 & +56 37 56.29 & 20.56 & 1.05,1.225,1.45,0.98 & 1.34  \\ \hline
3.14 & 37.1 & SDSSJ095454.73+373419.7 & 09 54 54.735 & +37 34 19.79 & 19.57 & 0.95,1.475,1.65,0.90 & 1.544 \\
     &      & SDSSJ095454.99+373419.9 & 09 54 54.999 & +37 34 19.99 & 18.91 & 1.45,1.575,1.95,0.94 & 1.892 \\ \hline
4.76 & 10.2 & SDSSJ114718.44+123439.8 & 11 47 18.448 & +12 34 39.84 & 20.91 & 1.45,1.625,2.00,0.66 & 1.583 \\
     &      & SDSSJ114718.66+123436.3 & 11 47 18.668 & +12 34 36.33 & 19.80 & 2.15,2.225,2.60,0.54 & 2.232 \\ \hline
3.06 &  5.6 & SDSSJ120450.54+442835.8 & 12 04 50.543 & +44 28 35.89 & 19.04 & 0.95,1.125,1.45,0.98 & 1.144 \\
     &      & SDSSJ120450.78+442834.2 & 12 04 50.784 & +44 28 34.25 & 19.48 & 1.35,1.725,1.95,0.78 & 1.814 \\ \hline
5.04 & 23.3 & SDSSJ124948.12+060709.0 & 12 49 48.127 & +06 07 09.04 & 20.41 & 2.20,2.325,2.65,0.89 & 2.001 \\
     &      & SDSSJ124948.17+060714.0 & 12 49 48.179 & +06 07 14.02 & 20.38 & 1.85,2.075,2.20,0.58 & 2.376 \\ \hline
3.01 & 20.0 & SDSSJ125530.44+630900.5 & 12 55 30.445 & +63 09 00.51 & 20.30 & 1.50,1.675,1.90,0.88 & 1.753 \\
     &      & SDSSJ125530.82+630902.0 & 12 55 30.823 & +63 09 02.09 & 20.60 & 1.10,1.375,1.50,0.98 & 1.393 \\ \hline
4.94 & 31.2 & SDSSJ142359.48+545250.8 & 14 23 59.484 & +54 52 50.83 & 18.63 & 1.00,1.175,1.45,0.973 & 1.409\tablenotemark{I} \\
     &      & SDSSJ142400.00+545248.7 & 14 24 00.006 & +54 52 48.79 & 19.93 & 1.45,1.575,1.90,0.772 & 0.610\tablenotemark{I} \\ \hline
4.35 & 25.2 & SDSSJ162902.59+372430.8 & 16 29 02.594 & +37 24 30.85 & 19.17 & 0.80,0.975,1.10,0.93 & 0.923 \\
     &      & SDSSJ162902.63+372435.1 & 16 29 02.634 & +37 24 35.17 & 19.35 & 0.70,0.925,1.10,0.98 & 0.906 \\ \hline
5.83 &  176 & SDSSJ171334.41+553050.3 & 17 13 34.414 & +55 30 50.36 & 18.88 & 1.00,1.375,1.45,0.968 & 1.276\tablenotemark{I} \\
     &	    & SDSSJ171335.03+553047.9 & 17 13 35.037 & +55 30 47.91 & 19.11 & 2.00,2.175,2.20,0.686 & star\tablenotemark{I} \\ \hline
5.06 &  172 & SDSSJ211157.24+091559.3 & 21 11 57.248 & +09 15 59.33 & 20.73 & 0.95,1.275,1.40,0.995 &       \\
     &      & SDSSJ211157.26+091554.2 & 21 11 57.269 & +09 15 54.28 & 19.83 & 1.00,1.325,1.35,0.983 & star\tablenotemark{S}  \\ \hline

\sidehead{\underline{Binary quasars}}
3.94 & 18.8 & SDSSJ0959+5449A & 09 59 07.060 & +54 49 08.09 & 20.60 & 1.90,2.025,2.15,0.59 & 1.956 \\
     &      & SDSSJ0959+5449B & 09 59 07.471 & +54 49 06.38 & 20.07 & 1.40,1.575,2.10,0.90 & 1.954 \\ \hline
3.55 &  6.5 & SDSSJ1259+1241A & 12 59 55.464 & +12 41 51.06 & 19.99 & 1.95,2.175,2.30,0.43 & 2.180 \\
     &      & SDSSJ1259+1241B & 12 59 55.617 & +12 41 53.81 & 20.09 & 1.90,2.175,2.25,0.52 & 2.189 \\ \hline
3.81 &  4.3 & SDSSJ1303+5100A & 13 03 26.144 & +51 00 51.00 & 20.54 & 1.50,2.075,2.20,0.82 & 1.686 \\
     &      & SDSSJ1303+5100B & 13 03 26.177 & +51 00 47.21 & 20.37 & 1.60,1.775,2.00,0.93 & 1.684 \\ \hline
3.12 &  0.5 & SDSSJ1337+6012A & 13 37 13.085 & +60 12 09.70 & 20.04 & 1.30,1.775,2.05,0.66 & 1.721 \\
     &      & SDSSJ1337+6012B & 13 37 13.133 & +60 12 06.60 & 18.59 & 1.50,1.625,1.95,0.94 & 1.727 \\ \hline
5.13 &  4.1 & SDSSJ1432-0106A & 14 32 28.949 & -01 06 13.55 & 21.10 & 1.55,2.125,2.25,0.69 & 2.082 \\
     &      & SDSSJ1432-0106B & 14 32 29.247 & -01 06 16.06 & 17.83 & 1.90,2.025,2.15,0.96 & 2.082 \\ \hline
4.11 &  7.2 & SDSSJ1530+5304A & 15 30 38.564 & +53 04 04.03 & 20.56 & 1.45,1.575,1.95,0.65 & 1.531 \\
     &      & SDSSJ1530+5304B & 15 30 38.824 & +53 04 00.65 & 20.70 & 1.40,1.725,2.15,0.93 & 1.533 \\ \hline
3.90 &  2.8 & SDSSJ1637+2636A & 16 37 00.881 & +26 36 13.71 & 20.61 & 0.45,0.575,0.85,0.46 & 1.961\tablenotemark{D}\\
     &      & SDSSJ1637+2636B & 16 37 00.932 & +26 36 09.87 & 19.36 & 1.40,1.525,1.80,0.64 & 1.961\tablenotemark{D}\\ \hline
3.72 &  3.9 & SDSSJ1723+5904A & 17 23 17.307 & +59 04 42.79 & 20.31 & 1.45,1.725,2.25,0.63 & 1.597 \\
     &      & SDSSJ1723+5904B & 17 23 17.421 & +59 04 46.41 & 18.88 & 1.55,1.725,1.90,0.94 & 1.604 \\ \hline
5.81 & 35.0 & SDSSJ2214+1326A & 22 14 26.792 & +13 26 52.38 & 20.64 & 1.55,2.025,2.20,0.85 & 1.995 \\
     &      & SDSSJ2214+1326B & 22 14 27.032 & +13 26 57.01 & 20.34 & 1.65,1.825,2.05,0.96 & 2.002 \\ \hline
\sidehead{\underline{Confirmed lenses}}
3.76 &  1.9 & SDSSJ1004+4112A & 10 04 34.800 & +41 12 39.29 & 18.64 & 1.55,1.725,2.00,0.94 & 1.734\tablenotemark{i}\\
     &      & SDSSJ1004+4112B & 10 04 34.917 & +41 12 42.81 & 19.04 & 1.55,1.725,2.15,0.79 & 1.734\tablenotemark{i}\\ \hline
3.04 & 46.4 & SDSSJ1206+4332A & 12 06 29.648 & +43 32 17.57 & 18.78 & 1.65,1.825,2.05,0.96 & 1.789\tablenotemark{o}\\
     &      & SDSSJ1206+4332B & 12 06 29.652 & +43 32 20.61 & 19.38 & 1.95,2.175,2.35,0.53 & 1.789\tablenotemark{o}\\ \hline

\enddata

\tablecomments{
Components of a binary are denoted A and B so that the position angle from A to B lies between $0^\circ$ and $180^\circ$. This convention differs from H06, from which we take identifications and redshifts, except for objects labeled $S$ (taken from the SDSS), $D$ (discovered by \citealt{Sra78}, confirmed as a possible lens by \citealt{Djo84}, and likely a binary instead, e.g., \citealt{Koc99,Pen99,Rus02}), i (part of the quad lens from \citealt{Ina03}), o \citep{Ogu05}, and I \citep{Ina07}. Both quasars SDSSJ162902.59+372430.8 and SDSSJ162902.63+372435.1 first appear in \citet{Mas00}. SDSSJ1004+4112A was discovered by \citet{Cao99}, and SDSSJ1432-0106B by \citet{Hew91}. We note that we mistakenly listed SDSSJ095454.73+373419.7 as lying at z=1.554 in M07b. $g$ is not corrected for Galactic extinction. {\em See Table~\ref{table:singles} for additional notes describing shared notation}.
}

\end{deluxetable}

\begin{deluxetable}{cc}
\tabletypesize{\scriptsize}
\tablecaption{Breakdown of ($3\arcsec\leq\Delta\theta<6\arcsec$) DR4 KDE quasar pairs\label{table:bkdown}}
\tablecolumns{2}
\tablewidth{0pt}
\tablehead{
\colhead{Category} & \colhead{Number of Confirmed Pairs}
}
\startdata

Total binary quasar candidates & 98  \\
Total now identified & 45 \\
Likely binary quasars & 19\\
Quasar pairs separated in redshift & 18   \\
Pairs containing $\geq1$ non-quasars &  6  \\
Pairs that are confirmed lenses &  2 \\

\enddata

\tablecomments{It is possible that a few more objects listed as ``Likely binary quasars" may turn out to be a lensed quasar when scrutinized at higher resolution.}
\end{deluxetable}

\begin{deluxetable}{crrrrllrc}
\tabletypesize{\scriptsize}
\tablecaption{Complete, statistical, clustering subsample \label{table:clus}}
\tablecolumns{9}
\tablewidth{0pt}
\tablehead{
\colhead{Name} & \colhead{$\Delta\theta ('')$}  & \colhead{$\chi_{color}^2$} & \colhead{$R_{prop}$} & \colhead{$R$} & \colhead{$z_A$} & \colhead{$z_B$} & \colhead{$|\Delta v_{\|}|$} & \colhead{Table}
}
\startdata

SDSSJ0959+5449 &  3.94 & 18.8 & 23.8 & 70.2 & 1.956  & 1.954 & 200 & (\ref{table:old})        \\
SDSSJ1320+3056 & *4.74 & 32.0 & 28.8 & 74.7 & 1.595  & 1.597 & 200 & (\ref{table:bin})        \\
SDSSJ1418+2441 & *4.50  &  2.8 & 20.9 & 32.8 & 0.572  & 0.573  & 100 & (\ref{table:bin})          \\
SDSSJ1426+0719 & 4.27  &  3.2 & 25.6 & 59.5 & 1.312  & 1.309  & 400 & (\ref{table:bin})          \\
SDSSJ1458+5448 & 5.14  & 65.6 & 31.1 & 90.2 & 1.913  & 1.912 & 0 & (\ref{table:bin})         \\
SDSSJ1507+2903 & 4.35  & 14.4 & 23.8 & 44.3 & 0.875? & 0.862? & 2100 & (\ref{table:ambig}) \\
SDSSJ1530+5304 & 4.11  &  7.2 & 24.9 & 63.1 & 1.531  & 1.533  &   200  & (\ref{table:old})           \\
SDSSJ1635+2911 & *4.92 & 25.6 & 29.9 & 77.3 & 1.582  & 1.590  &  900 & (\ref{table:bin})         \\

\enddata

\tablecomments{The DR4 KDE binary quasar candidate sample is now spectroscopically complete for component separations $3.9\arcsec < \Delta\theta < 5.2\arcsec$ for $g < 20.85$ in regions with Galactic absorption $A_g < 0.17$. 20 pairs meet these criteria, and 8 of them are (the listed) binary quasars. A * denotes a possible lens (see note in Table~\ref{table:bin}). $R_{prop}$ ($R$) is the transverse proper (comoving) separation ($h^{-1}~{\rm kpc}$). $|\Delta v_{\|}|$ is the line-of-sight velocity difference (${\rm km~s^{-1}}$). The final column ``Table'' denotes where we first listed these binaries. The 5 listed binaries with transverse separations of $23.7 \leq R_{prop} \leq 29.9$ represent a spatially complete subsample for redshifts of $1.03 < z < 2.10$.}
\end{deluxetable}

\end{document}